\documentclass[epj,nopacs,final]{svjour}

\usepackage{epsfig}
\usepackage{amssymb}
\usepackage{hyperref}

\newcommand{\etal}{\textit{et al.}}

\begin{document}

\title{Soft colour interactions and diffractive Higgs production\thanks{Talk presented by R.E.\ at HEP2003 Conference, Aachen, Germany.}}


\date{\today}

\author{R.~Enberg\inst{1}\thanks{Present address:\ Service de physique th\'eorique, CEA/Saclay, F-91191 Gif-sur-Yvette Cedex, France}
\and G.~Ingelman\inst{1,2} 
\and N.~T\^\i mneanu\inst{1}\thanks{Present address:\ Department of Cell and Molecular Biology, Uppsala University, 
Box 596, S-75124 Uppsala, Sweden}}
\institute{
High Energy Physics, Uppsala University, 
Box 535, S-75121 Uppsala, Sweden 
\and
Deutsches Elektronen-Synchrotron DESY, 
Notkestrasse 85, D-22603 Hamburg, Germany 
}

\abstract{The topical subject of Higgs production in diffractive hard scattering
events at the Tevatron and LHC is discussed. This has been proposed as
a Higgs discovery channel with appealing experimental features.
Predictions are obtained from the Soft Colour Interaction model, where
rapidity gaps are created by a new soft interaction added to the normal
hard scattering processes, implemented in the Monte Carlo event
generator PYTHIA. A brief review of the successful application of the
model to describe all CDF and D\O\ data on diffractive hard scattering,
such as production of $W/Z$, dijets, beauty and $J/\psi$ is also given.
}

\maketitle


\section{Introduction}
Recently there has been a lot of interest in searching for the Higgs boson at the Tevatron or LHC through diffractive hard scattering, see \cite{KMR,CR} for recent overviews with references. Predictions of cross sections from various models vary by orders of magnitude, however, and therefore it is  important to consider predictions of the same models for observables in other experimental channels, that can be compared to existing or coming data from the Tevatron.

Two such observables in events with two leading protons, so-called double pomeron exchange events (DPE), are dijet production at the Tevatron, and production of pairs of photons with large invariant mass at the Tevatron and LHC. There exist data for the former~\cite{CDF-DPE,AW} and data is expected for the latter.

In this talk we summarize our predictions from the soft colour interaction (SCI) model \cite{SCI} for diffractive Higgs boson production \cite{PRL,PRD2} and diphoton production \cite{PRD2} as well as the comparison of results for dijet production from CDF with our calculations \cite{PRD2,PRD}.

\section{The model}
The SCI model \cite{SCI} was developed in an attempt to better understand non-perturbative QCD dynamics and provide a unified
description of all final states. The basic assumption is that soft colour
exchanges give variations in the topology of the confining colour string-fields
which then hadronize into different final states, with and without
rapidity gaps or leading protons. 

\begin{figure}[t]
\begin{center}
\epsfig{width= 0.9\columnwidth,file=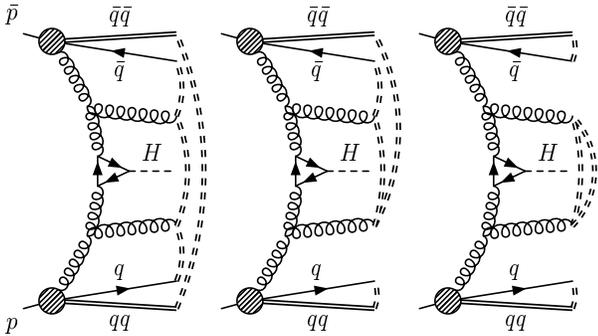,clip=}
\caption{Higgs production in $p\bar{p}$ collisions with string topologies
(double-dashed lines) before and after soft colour interactions in the SCI model,
resulting in events with one or two rapidity gaps (leading protons).}
\label{pp-higgs}
\end{center}
\end{figure}

To be able to use the SCI model for hadron-hadron collisions, it has been implemented in the Monte Carlo program PYTHIA \cite{Pythia}. The hard parton level interactions are given by
standard perturbative matrix elements and parton showers, which are not altered
by the softer non-perturbative effects. The SCI model then applies an explicit
mechanism where colour-anticolour (corresponding to non-perturbative gluons) can
be exchanged between the emerging partons and hadron remnants. The probability
for such an exchange cannot yet be calculated and is therefore taken to be a
constant given by a phenomenological parameter $P$. These colour exchanges
modify the colour connections between the partons and thereby the colour
string-field topology, as illustrated in Fig.~\ref{pp-higgs}. Standard Lund
model hadronization \cite{Lund} of the string fields then leads to different
final states, with gaps in rapidity regions where no string was present.
These soft processes do not affect
the cross section for the hard scattering process, but only the distribution of
hadrons in the final state. Diffractive events are
then selected using one of two criteria: (1) a leading (anti)proton with $x_F>0.9$
or (2) a rapidity gap in the forward or backward region, for example
$2.4<|\eta|<5.9$ as used by the CDF collaboration.

For a detailed description of the model and its application to diffractive hard scattering in hadron-hadron collisions, see Ref.\ \cite{PRD}.

\section{Results}

The SCI model has previously been shown to be very successful in reproducing diffractive HERA data, e.g., for the diffractive structure function and various final state observables \cite{SCI,MC99-SCI}. Furthermore, in an earlier paper \cite{PRD}, we showed that \emph{all} diffractive data from CDF and D\O{} existing at the time could be reproduced with the \emph{same} model. This is a  nontrivial result, and in fact there is no other model for diffraction in the literature that can achieve this without various types of modifications. This gives very strong support to the SCI model and to the reliability of our predictions for diffractive Higgs production.

Let us begin by considering single diffractive (SD) hard scattering. In ref.\ \cite{PRD}, we reproduced the diffractive ratios $R=\sigma^\mathrm{diff}/\sigma$ for SD production of $W$, $Z$, $b\bar b$, $J/\psi$ and dijets, as well as various kinematical distributions for dijets. For dijets, the CDF collaboration has also made measurements is DPE events, and we get a reasonable agreement with the cross section, and good agreement with the transverse momentum dependence of the cross section as well as the dijet mass fraction.

The latter is defined as the fraction of the total invariant mass of the system without the leading protons that the pair of jets takes, and is sensitive to the amount of extra radiation in the event. It is shown in Fig.\ 2 together with the CDF data~\cite{CDF-DPE}.

\begin{figure}[bhtp]
\begin{center}
\epsfig{width= 0.95\columnwidth,file=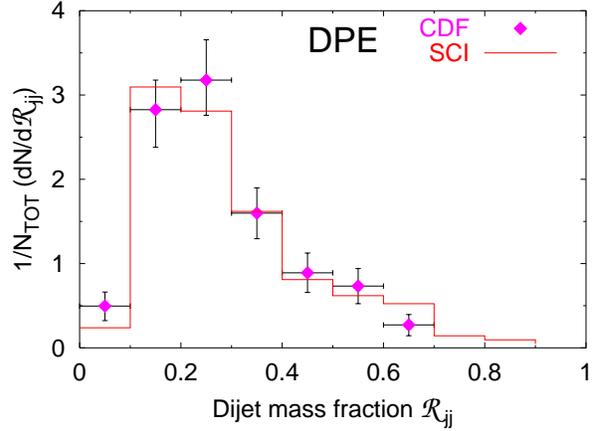}
\caption{Distribution of the dijet mass fraction, i.e., \ ratio ${\cal R}_{jj}$ of
the invariant mass of the dijet system to the invariant mass of the central
hadronic system in DPE events. CDF data \protect\cite{CDF-DPE} compared to the
SCI model.}
\label{massfraction}
\end{center}
\end{figure}

For so-called exclusive production of dijets, this ratio should be equal to one, although experimental smearing would give a peak at slightly lower mass fraction~\cite{AW}. This is clearly not seen in the data in fig.\ 2, nor in the newer data presented at this conference \cite{AW}. Our simulations, however, neatly reproduce the measured distribution.

Turning now to Higgs production, we select all hard subprocesses in PYTHIA
which  produce a Higgs boson. The dominant one is $gg\to H$ via a quark loop, which accounts for 50\% and 70\% of
the cross section, depending both on the Higgs mass and the
center of mass energy. Our results in the following are based solely on leading
order cross sections.

Applying the SCI model on the resulting partonic state gives rise to
different color string topologies, resulting in different final states after
the standard Lund model \cite{Lund} has been applied for hadronization
(Fig.~\ref{pp-higgs}). The results for diffractive Higgs production are
shown in detail in \cite{PRL} and we summarize the main results in terms of overall cross sections in Table~\ref{tab-higgs}. 

At the Tevatron, the cross section for Higgs in DPE events, which would have the
least disturbing underlying hadronic activity, is too small to give any
produced events. Higgs in single diffractive events are produced, but in small
numbers such that only the decay mode $H\to b\bar{b}$ with the largest
branching ratio will give any events to search for.
\begin{table}[t]
\caption{Cross sections at the Tevatron and LHC for
Higgs in single diffractive (SD) and DPE events, defined by leading protons or
rapidity gaps, obtained from the soft color interaction model (SCI).
\label{tab-higgs}}
\begin{tabular}{lllcc}
\hline
\hline
        & & & Tevatron & LHC \\
\multicolumn{3}{l}{$m_H=115$~GeV} & $\sqrt{s}=1.96$~TeV & $\sqrt{s}=14$~TeV \\
 \hline
Total &$\sigma [\mbox{fb}]$  & & {$600$}        & {$27000$} \\
\hline 
SD &$\sigma \; [\mbox{fb}]$ leading-p  & & $1.2$  & $190$  \\
   &$\sigma \; [\mbox{fb}]$ gap       &  & $2.4$ & $27$    \\
\hline 
DPE &$\sigma \; [\mbox{fb}]$ leading-p's& &  $1.2 \cdot 10^{-{4}}$  &  $0.19$  \\
    &$\sigma \; [\mbox{fb}]$ gaps      & & $2.4 \cdot 10^{-{3}}$  &  $2.7\cdot 10^{-{4}}$   \\
\hline
\hline
\end{tabular}
\end{table}

At LHC, the high energy and luminosity facilitates a study of single diffractive
Higgs production, where also the $H\to \gamma \gamma$ decay should
be observed. A few DPE Higgs events may be observed, but these events will not be as clean as naively expected. The available energy is enough to produce the Higgs and the leading protons as well as an underlying event that will populate forward detector rapidity regions with particles \cite{PRL}, and the rapidity gap will be in the very forward region not covered by detectors.
This causes a much lower diffractive cross section when requiring a gap instead
of a leading proton at LHC.

Finally, we want to discuss the production of a pair of photons, either as decay products of the Higgs boson, or a pair of prompt photons.
The $H\to\gamma\gamma$ decay mode is of experimental interest due to its clear
signature with two photons of high $p_\perp$. Its branching
ratio is, however, quite low since it proceeds via a higher order loop diagram.
Therefore, this
decay mode gives too low rates to be observable in diffractive interactions, except for a handful $H\to\gamma\gamma$ in single
diffraction at LHC. 
 
\begin{figure}[t]
\begin{center}
\epsfig{width= 1\columnwidth,file=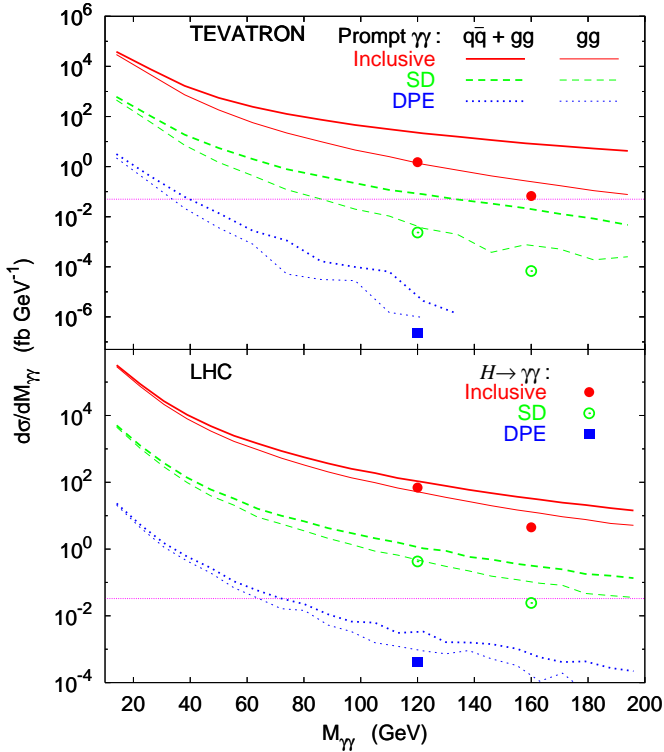}
\caption{Differential cross sections at the Tevatron and LHC for the production of prompt
$\gamma\gamma$ (with $|\eta_\gamma |<2$) as a function of diphoton invariant mass. Curves are predictions from the soft color
interaction model; inclusive, single diffractive and DPE events
showing separately the contribution from the $gg\to\gamma\gamma$ process (thin
lines). For comparison, $\sigma(H)\cdot{\rm BR}(H\to\gamma\gamma)$ is shown for $m_H=120$ and 160 GeV. The horizontal lines show the cross-section for obtaining one event with
the planned luminosity at the Tevatron and LHC.}
\label{fig-gammagamma}
\end{center}
\end{figure}

The requirement of the two
high-$p_\perp$ photons to be essentially back-to-back in the transverse plane
and isolated, removes the major backgrounds of photons as decay products in
jets. The serious  background to this signal is given by the production of a pair of prompt high-$p_\perp$ photons from the hard processes
$q\bar{q}\to\gamma\gamma$ and $gg\to\gamma\gamma$. In Ref.\ \cite{PRD2} we show that this background is always larger than the signal from Higgs decay and conclude that $\gamma\gamma$ is not a straightforward signal for observing the Higgs boson in diffractive events. 

On the other hand, prompt $\gamma\gamma$ production has a large enough cross
section to be investigated in diffractive events. 
At the Tevatron, observable rates are predicted for SD events with
$p_\perp^\gamma$ up to $\sim 75$ GeV, so the model can be tested against data
even up to scales of order $m_H/2$, but for DPE, $p_\perp^\gamma$ is only
observable up to 15-20 GeV.

Similarly at the LHC, prompt $\gamma\gamma$ gives observable rates for SD for
$p_\perp^\gamma$ up to $\sim 100$ GeV and for DPE up to $\sim 50$ GeV, i.e.\  up
to the scale of $m_H/2$. 

Thus, the basic mechanism for producing rapidity gap events with a high mass
state from $gg$ fusion via a quark loop can be tested already at the Tevatron
and further investigated at the LHC.

\section{Conclusions}
To summarize, we have shown results supporting the conclusion that the SCI model, though very simple, is able to reproduce a vast array of experimental data and is therefore likely to give trustworthy results also for the diffractive production of Higgs bosons. We have therefore provided predictions for diffractive production of Higgs, and also for dijets and diphotons, that can be used to test the model against even more data. It is worth mentioning again, that data from both diffractive DIS at HERA and from diffractive hard scattering in $p\bar p$ collisions at the Tevatron are correctly described. This is a unique feature of the SCI model, and further supports its use.

The predicted cross section for single diffractive Higgs production at the Tevatron is too low to be useful, and for DPE events, the cross section is far too small to yield an observable event rate. At the LHC both single diffractive and DPE events should be possible to observe for a Standard Model Higgs with a mass between 100 GeV and 200 GeV. However, diffractive events are not as clean as expected, with a lot of radiation in the forward regions even in events with leading protons. Similarly, the cross section for double prompt photons is very small at the Tevatron.

The phenomenological success of this simple model also makes it interesting to consider whether one could find a more theoretical basis for the  colour-carrying soft exchanges occurring in the model. Work in this direction is currently in progress \cite{BEHI}.

\end{document}